\documentstyle[preprint,aps,epsf]{revtex}
\begin{document}
\input psfig.tex
\draft
\title{Self - organized - criticality and synchronization in\\
pulse coupled relaxation oscillator systems; the\\
Olami, Feder and Christensen and the Feder\\
 and Feder model}
  \author{Samuele
Bottani, Bertrand Delamotte }
 
\address{Laboratoire de
Physique Th\'eorique et Hautes Energies, \\
 Universit\'{e} Pierre et Marie Curie --  Paris VI,  Universit\'{e} Denis Diderot--  Paris
VII\\ T24-5e, 2 place Jussieu, 75251 Paris Cedex 05,
France}
\date{}

\maketitle
 
\begin{abstract}
We reexamine the dynamics of the Olami, Feder and Christensen (OFC) model. We show that, depending on the dissipation,
it exhibits two different behaviors and that it can or cannot show  self - organized - criticality (SOC) 
and/or synchronization. We also show 
that while the Feder and
 Feder model perturbed by a stochastic noise is  SOC and has the same exponent for the distribution 
of avalanche sizes as the 
OFC model, it does not show synchronization. We conclude that a relaxation oscillator system can be 
synchronized and/or SOC and 
that therefore synchronization is not necessary for criticality in these models.
\\ 
LPTHE preprint 9605
\end{abstract}
\newpage
\section{Introduction}
The absence of characteristic scales in numerous natural phenomena has motivated the introduction of the
concept of Self-Organized-Criticality (SOC)\cite{BAKI}. According to this concept, scale  invariance would
be ``naturally" and dynamically generated in out of equilibrium systems. Up to now, it is not known if a few
basic mechanisms are at work in SOC or if each scale  invariant phenomenon is a particular case. The
dynamics of most of the known SOC models is an avalanche dynamics due, in many cases, to a threshold dynamics
at the microscopic level. For instance, in the sandpile model, once the sand height exceeds the threshold on
one site, it relaxes at that site and being redistributed on the neighbors, can trigger an avalanche of
relaxations. In this kind of models, the avalanches have no characteristic scale if the local state
variable, the sand in our example, is conserved. However, this conservation is not in general necessary for
criticality. In particular, the whole class of coupled relaxation oscillator systems exhibiting SOC do not
require conservation. In this case, the critical exponent $\nu$ of the avalanche distribution $P(s)\sim s^{-\nu}$
can depend continuously
on the dissipation.  In the following, we study two of these systems, the Olami, Feder and Christensen
(OFC) model and variants of the Feder and Feder (FF) model and we propose an explanation of their dynamics. 
In the OFC model, we show in particular that, depending on $\nu$, there are two qualitatively different 
regimes of the dynamics in the thermodynamical limit. For small $\nu$ (low dissipation) the dynamics is dominated by 
 large avalanches and is highly non trivial. On the contrary, for large $\nu$ (high dissipation), the dynamics is dominated by
small avalanches.
We show in this case that,  as $\nu$ increases, due to the {\it spatial distribution} of large avalanches, 
the dynamics becomes more and more trivial, {\em in the
thermodynamical limit} --- but on the boundaries ---, even if the multi-site avalanches are still distributed as power laws. 
In both cases, the system 
exhibits also synchronization, the stability of which depends on the dissipation. When the system is almost conservative, the 
synchronization is almost absent and, when $\alpha$ is small, it exhibits very stable synchronization.

\section{The model}

The definition of a lattice relaxation oscillator system is the following. The free, i.e. uncoupled,
evolution of the state variable $E_i$ of the oscillator at site $i$ is given by $E_i= f(\phi_i)$ where
$\phi_i$ is the phase of the oscillator $i$, $\phi_i = a t\ \mbox{\rm mod}\  1/a$ where $t$ is the time and $1/a$ the
period.  For convenience, we  call $E_i$ the local stress. $f$ is a continuously increasing function on
$[0,1]$ such that $f(0)=0$ and $f(1)=1=E_c$. $E_c$ is called the threshold. Thus $E_i$ increases up to the
threshold and then relaxes instantaneously to 0. For uncoupled oscillators, once $f$ is given, $E_i(t)$ is a
periodic function with  fixed frequency. These oscillators are usually called Integrate and Fire
oscillators. They are widely used in biology for the modelization of biological rhythms (see for instance
\cite{Glass}). In the following, $f$ is supposed linear except stated otherwise. The interaction between the
oscillators is chosen such that when a site relaxes to zero ($E_i\ge E_{c}$) it emits a pulse $\Delta_i$
that increments all its neighbors: 
\begin{equation}
\begin{array}{l} E_i\to 0\\
 E_j\to E_j + \Delta_i\ \ \ \ \ \ \  \mbox{\rm where $j=$ nearest neighbors of $i$}
\end{array}
\end{equation}
 The oscillator $j$ is therefore phase advanced and relaxes also to zero if it exceeds its threshold.
Therefore an avalanche of relaxations can occur that stops only when no more  oscillator is unstable.

Several studies of systems of pulse coupled Integrate and Fire oscillators on a lattice have recently been
performed and a large variety of behaviors have been observed depending on the boundary
conditions\cite{GrassbergerI,Socolar,Middleton}, on the shape of the function $E_i(\phi)$\cite{Corral} and on the
choice of
 the coupling $\Delta_i$\cite{OlamiI,FederI,ChristensenI,Herz2}. In this article, we study the dynamics of
the two dimensional OFC model\cite{OlamiI,ChristensenI,Christensen3,Christensen4,ChristensenII} which is a
discretized and simplified version of the Burridge-Knopoff model of earthquakes\cite{Burridge}. It
 is  a model of pulse coupled Integrate and Fire oscillators with an especially rich
 dynamics. In this model the pulse $\Delta_i$ is proportional to $E_i$ just before the relaxation: $\Delta_i
=\alpha E_i, E_i\ge E_c$ and $\alpha\le1/4$ is a parameter that takes care of the dissipation. This model
was shown to display SOC for a wide range of parameters
$\alpha$\cite{OlamiI,ChristensenI,Christensen4,ChristensenII}. It has also a complex spatio - temporal
behavior with clustering of almost periodic sequences of large avalanches and a multi - fractal distribution in
time of the
 avalanche sizes\cite{Christensen3}. 
Let us recall the principal results on the OFC model relevant for our study.
The largest simulations have been performed by Grassberger\cite{GrassbergerI} and Middleton and Tang\cite{Middleton}
who have found power law distributions for the avalanche sizes for a wide range of $\alpha$, including
values as small as $0.07$\cite{Middleton} and $0.05$\cite{GrassbergerI}. For sufficiently large $\alpha$ 
the SOC behavior is well established and confirmed  by the finite size scaling. For small $\alpha$, the conclusions
are not so firmly established since the cut-off of the distribution does not obey the correct finite size scaling.
This is probably due to the transients that are extremely long and forbid to be in the asymptotic regime\cite{GrassbergerI}.
Grassberger has also suggested that there exists a critical value $\alpha_c$ of $\alpha$, $\alpha_c\sim 0.18$,
below which  the interior of the system would
almost be decoupled from the boundaries. According to this author, the bulk would then behave as in the case of periodic boundary conditions, where all the
avalanches are of size one and occur perfectly periodically. 
Grassberger also noticed that
most of the multi-site avalanches are triggered near the boundaries of the
system. 

\begin{figure}
\hspace{7cm}
\centerline{{\epsfxsize=7cm \epsffile{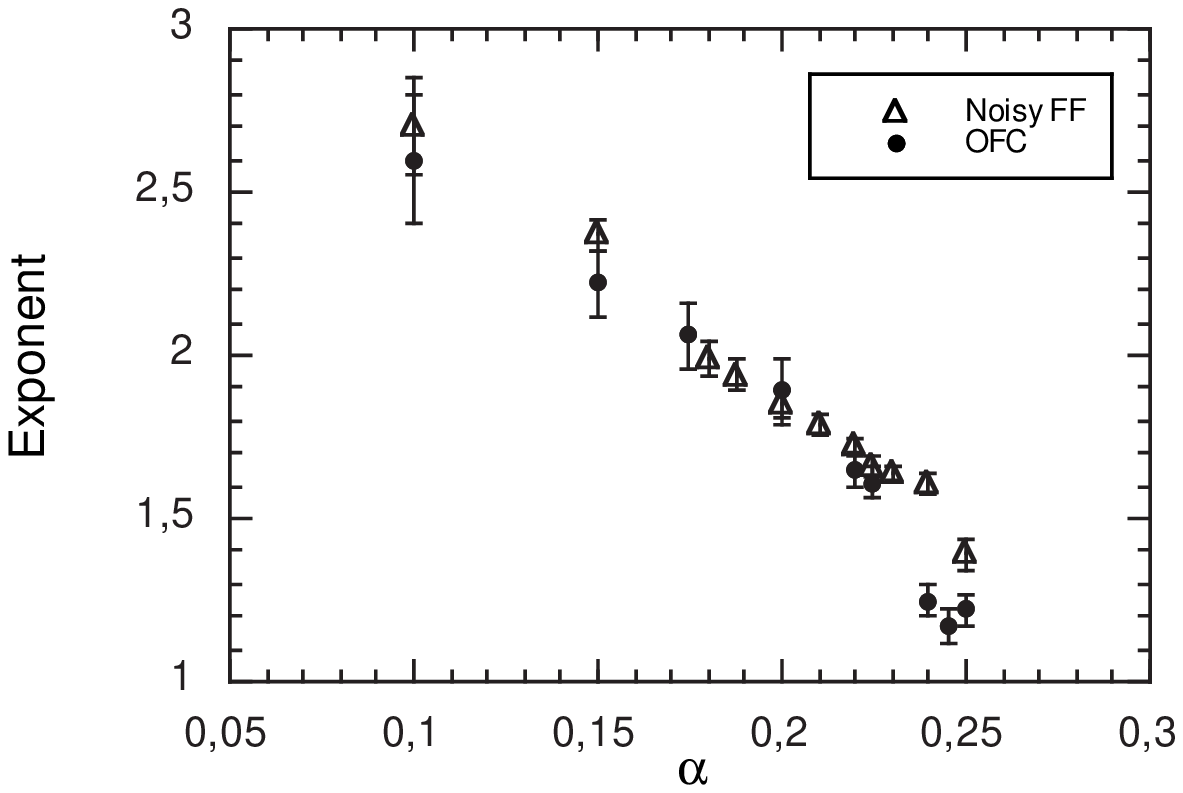}}}
\caption[Exposants critiques du modle d'Olami, Feder et Christensen et du  modle Feder 
et Feder en fonction de la dissipation]{\renewcommand{\baselinestretch}{1}\sl Plot of the exponents 
$\nu$ of the distributions of avalanche sizes $P(s)\sim s^{-\nu}$ in the FF
and OFC models for different values of $\alpha$. Up to $\alpha=0.23$, the noisy FF and OFC models have the
same exponents. For $\nu=2$, $\alpha$ is roughly $0.18$. }
\label{exponentofcff}
\end{figure}

In the following, we shall confirm the existence of two domains of $\alpha$ corresponding to two 
qualitatively different behaviors of the system. By using a very crude approximation,
we shall relate the existence of these two regimes to the existence of a critical value $\nu_c$ of the exponent 
$\nu$ of the distribution $P(s)$,
separating the  domain $(\nu<\nu_c)$, where, {\em statistically},
the large avalanches penetrate in the bulk, from the  domain, $(\nu>\nu_c)$, where they do not.
In our approximation, $\nu_c=2$, a value for which $\alpha$ is  indeed close to $0.18$, see Fig.\ref{exponentofcff}.

Christensen proposed that the behavior of this model arises from a tendency of the oscillators to
synchronize their relaxations\cite{ChristensenI}. Systems of Integrate and Fire oscillators can indeed evolve
towards large scale synchronization as shown in models with a global, all to all,
coupling\cite{MirolloII,ChristensenI,BottaniI,BottaniIII} and in lattice models with a strongly convex
function $E(\phi)$\cite{Corral,CorralI}. The question of the relationship between SOC and synchronization in
the OFC model was already  
 addressed by Middleton and Tang who studied the dynamics of this model for $\alpha =0.07$ and $\alpha =
0.15$ and who argued on theoretical grounds that there should exist a close relationship between these two
forms of organization\cite{Middleton}. However, to the best of our
 knowledge there has been no report of the direct observation of synchronization in the OFC model nor on its
possible relevance for the criticality of the OFC model. In this article, we reexamine the question
of this relationship between SOC and synchronization and propose a ``phase diagram" for all values of
$\alpha$ of the OFC model.

\section{The role of the spatial distribution of avalanches}

\begin{figure}
\centerline{{\epsfxsize=7cm \epsffile{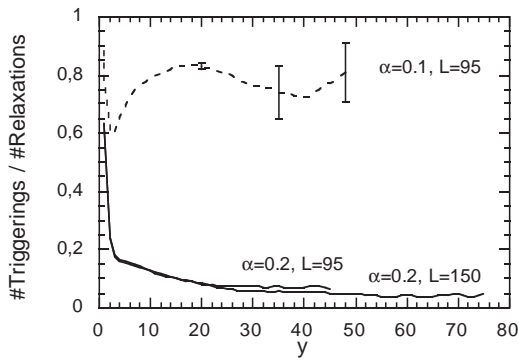}}}
\caption[]{\renewcommand{\baselinestretch}{1}\sl  Ratio between the number of times a site triggers an avalanche
and the number of times it relaxes as a function of the distance $y$ of the site to the nearest border.
When this ratio is one, all the avalanches at that site are of size one. When it is small, most of the avalanches are large. For
$\alpha=0.2$, the error bars are  small. For $\alpha=0.1$, the error bars increase with
$y$ since, in the center of the lattice, the number of sites used for averaging is far smaller that near the boundaries.
For $\alpha=0.1$, most avalanches in the bulk are of size one. For $\alpha=0.2$ on the contrary, 
the avalanches are large in the center of the lattice, independently of the lattice size.
 }
\label{Triggerings}
\end{figure}

\begin{figure}
\centerline{{\epsfxsize=7cm \epsffile{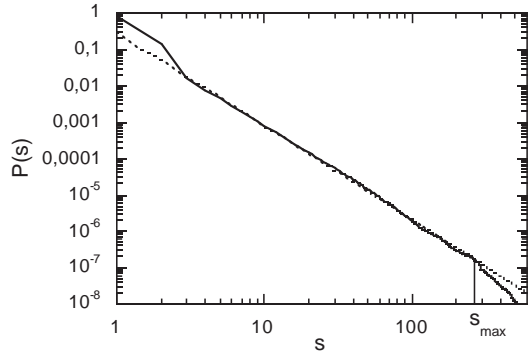}}}
\caption[]{\renewcommand{\baselinestretch}{1}\sl  Distribution $P(s)$ of avalanche sizes for $L=95$ and $\alpha=0.1$.
 The excess of  avalanches of size 1 and 2 is clearly seen. 
The dotted line is the power law fit $s^{-2.5}$. $s_{\mbox{\rm\scriptsize max}}$ is 
the size above which $P(s)$ is no longer fitted by a power law. }
\label{Distribution}
\end{figure}

In our opinion, the role of the spatial distribution of avalanches -- that is their localisation as a 
function of their size --
 has been underestimated in SOC models.
Let us now show why and why it is likely to be crucial for the OFC model. For all values of $\alpha$, 
it has been observed that 
``large'' avalanches are preferentially triggered near the boundaries\cite{GrassbergerI}, see Fig.\ref{Triggerings}. 
For tractability, we make the simplifying
 approximation, capturing the essential features of the model,
that they are {\em all} triggered on the boundaries of the system. Let us now show, by calculating 
the  percentage of large avalanches,
$s\in[{\bar s}(p),s\mbox{\rm\scriptsize max}]$, accounting for a percentage $p$ of the relaxations of a site in the bulk,
 that the system exhibits two different regimes depending on $\nu$.
For computational simplicity, we assume a sharp cut-off of the distribution $P(s)$ at the value 
$s_{\mbox{\rm\scriptsize max}}$ where the power law 
is no longer valid, see Fig.\ref{Distribution}. We thus neglect the tail of the largest avalanches, 
an hypothesis that we show is correct, 
at least for $\alpha$ small.

The scaling of 
$s_{\mbox{\rm\scriptsize max}}$ with the system size $L$ is not exactly 
known\cite{ChristensenII,GrassbergerI,Klein,Christensenqqch,Janosi}.
In the numerical simulations available, it seems that $s_{\mbox{\rm\scriptsize max}}$  scales for $\alpha>0.18$
 slightly faster than $L^2$\cite{ChristensenII}. This is
of course impossible, in the limit of large $L$, since, as we have verified numerically,
 there is no multiple relaxations of the oscillators 
inside an avalanche for $\alpha\stackrel{<}{\sim} 0.24$, i.e. for non conservative systems (see also \cite{Christensenqqch}).
As indicated in \cite{GrassbergerI,Christensenqqch}, this means that the true asymptotics is
not exactly reached with the system sizes studied.  We suppose in the following that the true asymptotic
behavior is given by $s_{\mbox{\rm\scriptsize max}}\sim L^\rho$. For $0.15<\alpha\stackrel{<}{\sim} 0.24$,  
the most reasonable assumption \cite{Janosi} is $\rho\simeq 2$.
For   smaller $\alpha$,  the determination of $\rho$ is extremely difficult due to very long transients.  We shall see however, that
our results are independent of its  precise value.

With our approximation, an elemenatry 
calculation  shows that
\begin{equation}
{\bar s}(p)=[s_{\mbox{\rm\scriptsize max}}^{2-\nu}(1-p)+p]^{1/(2-\nu)}
\end{equation}
Therefore, there are two different cases depending on $\nu$. If $1<\nu<2$, then ${\bar s}(p) \sim s_{\mbox{\rm\scriptsize max}}$ and the 
fraction of avalanches accounting for the percentage $p$ of relaxations goes to zero as $s_{\mbox{\rm\scriptsize max}}^{1-\nu}$ and therefore 
as $L^{\rho(1-\nu)}$. This means that the dynamics is completely dominated by large avalanches and that, in the thermodynamical limit,
a vanishing percentage of very large avalanches is responsible for a finite fraction of the relaxations. For $\rho=2$, these
 avalanches 
sweep  a finite fraction of the whole lattice, and therefore contribute to a finite fraction of the relaxations 
of the sites in the bulk.
Therefore, in this case, even if
the avalanches are triggered near the boundaries, we expect the dynamics of the bulk to be  non trivial, even in the
thermodynamical limit, since the avalanches are
 able to penetrate  into the system.  If true, this would
mean that the critical behavior of this system would be highly non -
 trivial since it would depend, even in the limit $L\to \infty$, on the boundary conditions
 through the avalanches triggered near the boundaries. 
Although the true spatial distribution of avalanches is more complicated than what we have supposed, and  $\rho$ could be slightly different from $2$, we  have 
been able to verify numerically, at least for $\alpha\ge 0.2$ and for lattice sizes up to $L=150$, that the relaxations in
 the bulk are mostly due to the  largest avalanches. Let us finally notice, that for general SOC models with $\nu<2$, 
the tails of the distributions $P(s)$  could actually dominate the dynamics, see however\cite{Christensenqqch}. Our conclusion for the previous case $1<\nu<2$ would however  be unchanged 
if we had taken into account this tail.

On the contrary, if $\nu>2$, then, in the limit $L\to \infty$, ${\bar s}(p)\sim p^{1\over 2-\nu}$ independently of $L$.
This means that the dynamics is dominated by small avalanches. In a first step, let us  suppose that the avalanches are 
compact, two dimensional objects. The avalanches that are able to reach a site in the bulk
have, in this case, a size $s$ of the order of $L^2$, $s\in [\eta L^2, s_{\mbox{\rm\scriptsize max}}]$ with $\eta$ less than one
and of the same order.
Since the dynamics is dominated by the small avalanches, the average number of relaxations per site, during $N$ avalanches, 
 scales as $N/L^2$.
Then the fraction $f$ of relaxations of a site in the bulk, due only  to large avalanches, behaves as:
\begin{equation}
f\sim {s_{\mbox{\rm\scriptsize max}}^{1-\nu}- (\eta L^2)^{1-\nu}\over 1/L^2}\sim L^{\rho(2-\nu)}.
\label{comportement1}
\end{equation}
 Therefore, we conclude that for $\nu>2$ and with our hypothesis,  a site in the bulk relaxes almost 
never in a large avalanche when $L\to\infty$. It is not difficult to show that this result is unchanged 
if the avalanches are one 
dimensional objects starting on the boundaries. In this case, one finds  in the most favorable case for the
penetration of avalanches in the bulk, $\rho=2$, that $f$ behaves as:
\begin{equation}
f\sim L^{2-\nu}.
\label{comportement2}
\end{equation}
In this case also, the avalanches are  not able to penetrate into the system and the conclusions are 
the same as  before. 

\begin{figure}
\centerline{{\epsfxsize=7cm \epsffile{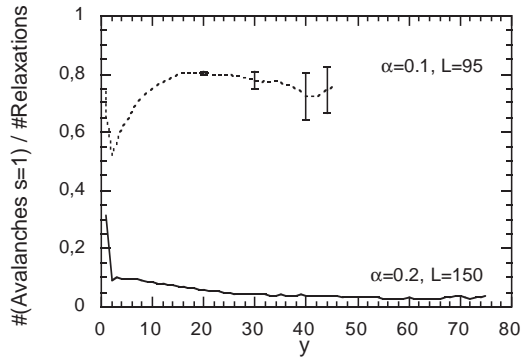}}}
\caption[]{\renewcommand{\baselinestretch}{1}\sl Plot of the percentage of relaxations
of a site, due to avalanches of size one, as a function of the distance $y$ of this site 
from the nearest border. The data were obtained with $10^{10}$ and  $14.10^6$ avalanches for
respectively $\alpha=0.1$ and  $\alpha=0.2$. 
A pure SOC behavior  for $\alpha=0.1$ (power  law fit of Fig.2) would predict an average of $37\%$ of relaxations in avalanches of size one.}
\label{Avalanche1}
\end{figure}

As  shown numerically for $\alpha=0.07$ and $\alpha=0.15$ in \cite{Middleton},
the toppling rate $r$ of any site in the bulk of the system behaves as a function of the distance $y$ from the boundary as:
\begin{equation}
r(y)= (1-4\alpha)^{-1} + c y^{-\eta}  
\label{mt}
\end{equation} 
where $c$ is a constant and $\eta$ an exponent that depends on $\alpha$. The relaxation rate is therefore 
almost constant,  at the thermodynamical limit, for a site in the bulk. We now show that this result combined 
with Eq.(\ref{comportement1}) or (\ref{comportement2}) is incompatible with a power law distribution of avalanche sizes 
for small avalanches.  Provided $\nu>2$, it is easy to show that,   whatever   the spatial distribution of the avalanches,
the number $N$ of avalanches necessary for all the sites in a band on the boundaries of finite  width $d$ to relax
once, scales as $N\sim L$. 
On the other hand, since a site in the bulk has almost the same relaxation rate as a site at distance $d$ 
from the boundary, for $d$ large enough, it 
relaxes also once during these $N$ avalanches. This is not possible if we assume 
a power law distribution of avalanche sizes valid from $s=1$ to $s=s_{\mbox{\rm\scriptsize max}}$, since the number of sites 
in the bulk that relax in small avalanches, i.e. mostly in avalanches of size 1, varies as $L^2$. We expect therefore 
a large excess of avalanches of size 1, a smaller of avalanches of size 2, etc\dots  This has been observed by Grassberger
for $\alpha = 0.05$ and we have verified   numerically for $\alpha=0.1,\  L=95,130$, with  
$10^9$ avalanches, where the system is in the steady state, that this is indeed correct.
In fact,  all the avalanches 
in the bulk are not of size 
one and it happens (rarely) that large avalanches are also triggered in the bulk. However, it is clear on 
Fig.\ref{Distribution} that the
 distribution $P(s)$ has indeed an excess of avalanches of 
size one and two and that the power law  really begins for $s\ge 3$. The fact that only two points are not 
aligned on the Log-Log curve should not be underestimated since the dynamics is precisely dominated by these small
avalanches for $\nu> 2$.  Moreover, Grassberger has also observed for $\alpha=0.05$, that
the relative weight of avalanches of size one increases with $L$.  For $\alpha=0.1$, $L=95$ and $10^{10}$ avalanches, the 
pure power law fit of $P(s)$ shown in Fig.\ref{Distribution} predicts that $37\%$ of the 
relaxations should be due to avalanches of size one, while our simulations show that
$75\pm 5\%$ of the relaxations  in the bulk are due to avalanches of size one, see Fig.\ref{Avalanche1}. 
This is confirmed by a simulation on a system 
of size $L=130$, where it is however more difficult to obtain a good statistics. When $\nu$ increases, 
i.e. $\alpha$ decreases, we expect that 
the relative weight of the avalanches of size one will get bigger and bigger.  
We conclude that, although the system seems to exhibit a SOC behavior, 
the boundaries and  the bulk are largely decoupled for $\nu>2$.
The fact that $s=1$ and $s=2$ do not fit the power law, unfortunately turns out to be crucial 
and spoils the critical behavior
in the thermodynamical limit. We have confirmed  this decoupling numerically for $\alpha=0.1$ 
and our arguments strongly suggest that it
is not a transient effect.

 Let us now emphasize that, within our approximation, the critical value of $\nu$ that 
separates the two regimes of the model is $\nu=2$, to which corresponds $\alpha\simeq 0.18$, see Fig.\ref{exponentofcff}. 
While it seems numerically that this value of $\alpha_c$ is reasonable, it is of course very difficult to conclude 
definitely about its value since the 
spatial distribution of avalanches is certainly non trivial. For $\alpha=0.15$, which is smaller but close to
$0.18$, it is numerically difficult, since it requires very large lattices,  to see the decoupling of the
interior of the system. 
It is, in fact, even difficult to predict if there exists a
sharp separation between two regimes. However, we think that our analysis clearly shows  that the existence of a
 power law distribution of the avalanche sizes, which is only a global information on the system, is not sufficient to
make sure that the physics is really scale invariant. The point that makes this issue non trivial, is that systems like the 
OFC model are not obviously translationally invariant, since their physics can  depend crucially on their boundaries, even 
at the thermodynamical limit. Moreover, we have shown that the whole behavior of a model, which is thought to be SOC, 
can actually  be dramatically affected by the fact that the power law distribution $P(s)$ is not verified 
for only one avalanche size, as it is the case in  the OFC model for small $\alpha$ and for $s=1$. 

\section{Synchronization}
Let us define more precisely what we mean by synchronization since
different behaviors are called synchronized in the
 litterature. Our definition of synchronization is the following: a cluster of oscillators is said to be perfectly 
synchronized when all the oscillators of the
cluster always relax together in the same avalanche. 
 We  say that a lattice is partially synchronized when  the whole system is not synchronized,
but contains clusters of sites that are synchronized. We shall also be interested in unstable
synchronization where the synchronized clusters have a large, but finite, life - time. Note that perfect
synchronization is of course incompatible with SOC. 
It is also important to be aware that our definition of synchronization does not imply that the
state variables of synchronized  oscillators  are degenerate. This is the case for  the
OFC and FF models where synchronized oscillators are in general non degenerate. 
Furthermore, synchronization should not be confused with a periodic activity of the lattice,
as the one observed for periodic boundary conditions, where the avalanches are all of size one and the period is 
$L\times L$\cite{Socolar,GrassbergerI}.
Since we are interested in 
the dynamics  of  avalanches,
we believe that it is our definition of synchronization which is physically relevant.

Let us call cycle the time necessary
 for the oscillators of the system to relax once. This notion is well - defined for the
 sites in the interior of the system since they all relax the same number of times in the same time interval, apart
from small fluctuations. We have studied by direct inspection and cycle
 after cycle, the shapes of the clusters of sites that relax simultaneously. For a wide range of parameters
$\alpha$, we have found partial and unstable synchronization of all length scales in the OFC model. To the
best of our knowledge, this is the first time that the presence of
 synchronization and its co-existence with  SOC is reported in the OFC model. Let us examine in greater
details this behavior when $\alpha$ varies. For $\alpha$ very small the oscillators are almost decoupled so
that most of the avalanches are of size one. The
 largest avalanches, that occur near the borders, are rare and small.
 The bulk behaves very much like the state described by Grassberger \cite{GrassbergerI}
for periodic boundary conditions and $\alpha<0.18$. In this state, called ``ordered'' by this author, the avalanches 
are only of size one and occur periodically one after the other. Since, in the bulk, there are no multi-site avalanches,
there is evidently no synchronization.

 As $\alpha$ increases in the interval $[0,0.18]$, larger avalanches occur
and synchronization begins to make sense. We have
 followed cycle after cycle the avalanches to which belongs a given site that has been previously chosen
randomly in the bulk. For $\alpha=0.1$, for instance, most of the avalanches in the bulk are of size one or two.
However, it happens that large avalanches -- large for this $\alpha$ means typically $s\ge 50$ -- are triggered 
in the interior of the system
and are extremely stable. Although we do not have statistics, we have been able 
to see on each synchronized cluster that, cycle after cycle,  avalanches are always triggered by the same site
and always for the same phase of this  site,
and that their life times are ususally of hundreds and even of  thousands of cycles. 
 We have checked that these results are independent of the test site chosen in the bulk.
 For $\alpha=0.15$, the
shape of these avalanches are also very stable with, however,  smaller life times of tenths of cycles. 
During these avalanches, only a few oscillators on the
border of the avalanches desynchronize. The
 synchronized clusters evolve  either by merging with a neighboring synchronized cluster or by breaking into
two clusters. These new clusters are themselves synchronized during many cycles until they  evolve by
breaking or merging. The system appears as a collection of synchronized clusters together with many sites that topple 
in avalanches of size one or two. The evolution of these clusters is the
superimposition of a mechanism of merging and breaking and a very slow evolution of the shape of these
clusters. Thus, for a given site, the nearest sites around it can belong to a synchronized subcluster that
does not break during hundreds of cycles. 

\begin{figure}
\centerline{{\epsfxsize=7cm \epsffile{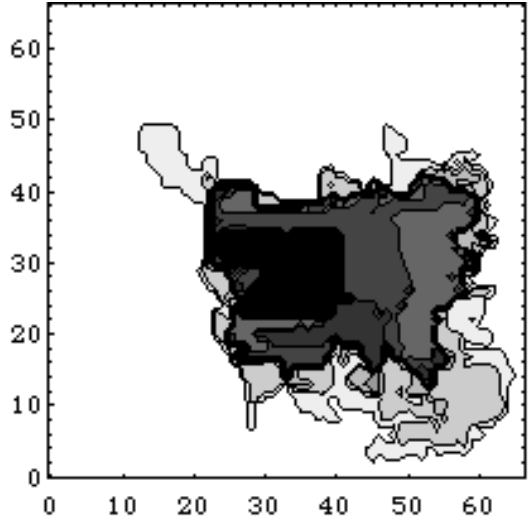}}}
\caption[Carte des oscillateurs ayant ŽtŽ synchronisŽs avec un oscillateur
 test]{\renewcommand{\baselinestretch}{1}\sl Map of  60 successive avalanches  involving the site
 of coordinates (30,30) for $\alpha=0.2$ and
$L=65$. The  grey level is an indication of the number of relaxations inside these avalanches. The blackest
spot corresponds to a cluster of sites that have always relaxed together during this sequence.}
\label{mapsynchro}
\end{figure}

When $\alpha$
 becomes larger than $0.18$, but still lower than $1/4$, say $\alpha \sim 0.2$, the system shows SOC together with
partial and unstable synchronization, see  Fig.\ref{mapsynchro}.
The synchronized clusters are much
  larger, in the average, than for $\alpha\le 0.18$, but evolve more rapidly. As seen on Fig.\ref{mapsynchro}, 
  a given site and its neighbors can belong to a synchronized subcluster that
does not break during hundreds of cycles. 
 We have observed that
synchronization occurs on all length scales, from small clusters of a few sites near the borders to
clusters  that represent a macroscopic fraction of
 the whole system. By following one particular site $i_0$ in the interior of the system,
 it appears that the neighboring sites are in general almost perfectly synchronized with $i_0$ and that the
level of synchronization decreases with the distance from $i_0$. Finally, when $\alpha$ becomes close to
1/4, say $\alpha=0.24$, the synchronization
 disappears almost completely.

Let us now try to understand the behavior of this system. As we argue in the following, its dynamics
involves a competition between a tendency towards synchronization, due to the open boundary conditions,
randomness and dissipation. Let us first clarify what we mean by randomness in the OFC model. The
microscopic rules of this model are entirely deterministic but randomness is anyway present because of the
initial conditions, where the phases are taken random. This randomness is maintained in the system since the
pulses $\Delta_i$ are proportional to
 $E_i$: $\Delta_i= \alpha E_i$. It is, of course, partially dissipated when $\alpha$ is less than 1/4  but, as
we shall see, the threshold dynamics is able to amplify the effect of a very small noise. 

Let us  study the synchronization in the OFC model. As shown by Socolar {\sl et al.}\cite{Socolar} and 
Middleton and Tang\cite{Middleton}, two isolated
oscillators of different frequencies, interacting with each other as in the OFC model, synchronize necessarily
with a frequency equal to the frequency of the slowest oscillator.  These authors argued that
since the oscillators on the boundaries receive less
 pulses by unit time,  since they have only three neighbors, than those far from the borders, their
effective frequencies should be lower than those in the interior. According to Middleton and Tang, this should
lead to a tendency towards synchronization and should be responsible for the large correlations in the OFC
model. However, as we have mentioned previously, the system does not always show SOC nor synchronization. Moreover,
its behavior depends strongly on $\alpha$, so that it is not clear whether and how this can produce large scale
synchronization and whether this is related to criticality.

\begin{figure}
\centerline{{\epsfxsize=7cm \epsffile{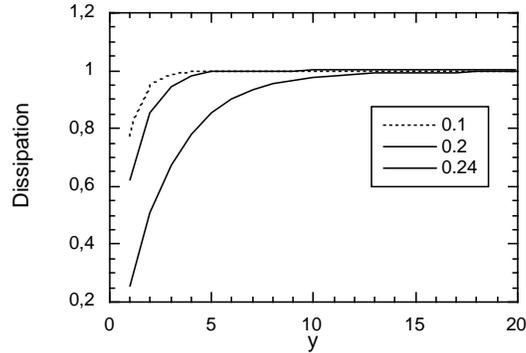}}}
\caption[Dissipation locale de chaque oscillateur]{\renewcommand{\baselinestretch}{1}\sl The local dissipation rate
 -- on each site -- as a function of the distance of the site to the border
of the system. Three values of $\alpha$ have been studied: $\alpha=0.1, 0.2, 0.24$. For a sufficiently large
distance, the dissipation is always 1.}
\label{dissipation}
\end{figure}

\begin{figure}
\centerline{{\epsfxsize=7cm \epsffile{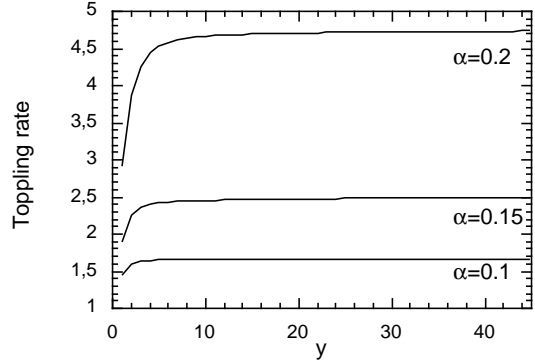}}}
\caption[Taux de relaxation en fonction de la distance aux bords]{\renewcommand{\baselinestretch}{1}\sl Relaxation 
rate $r_i$ of the oscillator $i$ as a function of the distance of $i$ to the border.
For $\alpha=0.1$ and $\alpha=0.15$, $r_i=(1-4\alpha)^{-1}$ in the bulk while for $\alpha=0.2$, $r_i$ is
slightly less than $(1-4\alpha)^{-1}$.}
\label{frequency}
\end{figure}

 As a first step, let us show that for $\alpha$
sufficiently large, the model cannot show stable  synchronization. In the following, we assume that for
$\alpha<1/4$, the system, which is highly dissipative, is such that, in the bulk, the system dissipates in the
average and on each site, $100\%$ of the stress it receives. Put it differently, there is no transport of stress from
the bulk to the boundaries. This assumption can be justified in the following way. In a subsystem of volume
$l\times l$, representing the interior of the system of volume $L\times L$, $l$ being of the order of $L$ for
large systems, $l\sim L$, the stress to be dissipated during a fixed time $t$ grows as $l^2$.  If a fraction
of this stress was dissipated on the boundaries, it would topple typically $L-l\propto L$ times from site to site, 
in large avalanches,  to go from the interior to the boundaries. Thus, only a fraction of this stress of
order  $(1-4\alpha)^{L-l}\sim (1-4\alpha)^{L}$ would arrive on the boundaries. 
Therefore, since $l^2(1-4\alpha)^L\to 0$ as $L$ increases, nothing of the interior
of the system can be dissipated on the boundaries for $\alpha<1/4$.
 In this sense, the conservative model $\alpha=1/4$ -- and possibly $\alpha$ very close to 1/4 -- is a very
particular case since everything is dissipated on the boundaries in this case.  We have verified numerically
that for any $\alpha$ not very close to $1/4$ (it would require extremely large lattices to study this
case), the dissipation rate is indeed $100\%$ on each site of the interior \cite{Middleton}, see Fig.\ref{dissipation}. 
This is expressed for a
site $i$ in the interior of the system by:
\begin{equation} 
{\bar E}_i(1- 4\alpha) r_i = 100\%=1
\label{dissipationofc}
\end{equation} 
with ${\bar E}_i$ being in this equation the average stress of the oscillator $i$ just before
the relaxation and $r_i$ the  relaxation rate of this oscillator. Now, if we suppose that a (large) 
cluster  of sites is permanently synchronized with a site on (or close to) the boundary, the relaxation rate  of which is
roughly $(1-3\alpha)^{-1}$  since it has three neighbors, see Fig.\ref{frequency}, we deduce  
from (\ref{dissipationofc}) that
\begin{equation} 
{\bar E}_i \simeq {1 - 3\alpha\over 1- 4\alpha}
\label{stressofc}
\end{equation} 
This is clearly impossible for $1- 4\alpha\to 0$,  since ${\bar E}_i $ diverges in this limit.
This means that the synchronization cannot be stable for large $\alpha$ and this comes from the fact that it
becomes harder and harder to synchronize oscillators as their ``natural"  frequencies become more and more
different. In fact, we can obtain a rough estimate of the value of $\alpha$ above
which the synchronization must be unstable. Since the propagation of the avalanche of relaxations occurs as
the propagation of a front, there is in the average two back - firings from the sites that relax to the sites
that have just relaxed before them in the avalanche. Thus, the value of a site after an avalanche is roughly
$2\alpha {\bar E}_i$.
 This value cannot be larger than one otherwise the back - firing would trigger another
relaxation. We have verified numerically that indeed, for generic values of
$\alpha$, not very close to $1/4$, there is never multiple relaxations of a site inside an avalanche
\cite{note1}.
Therefore, we deduce that stable
synchronization requires necessarily
\begin{equation}
 {1 - 3\alpha\over 1- 4\alpha}<  {1\over 2\alpha} \ \ \ \Longrightarrow\ \ \ \alpha<0.21
\label{borne}
\end{equation} 
The conditions (\ref{stressofc}) and (\ref{borne}) are of course  only necessary for a stable
synchronization, they are absolutely not sufficient. In particular, synchronization needs a very well
defined and stable spatial repartition of the values of the ${\bar E}_i$. For this reason, we expect
randomness to play a crucial role in the OFC model since randomness is never completely dissipated in this
model and since it tends to destabilize synchronized clusters. Randomness  is thus surely in competition with
synchronization\cite{note2}. In fact, we can expect that since ${\bar E}_i$ is bounded, the relaxation rate $r_i$ behaves
roughly as $(1-4\alpha)^{-1}$  in the interior when $\alpha$ is not too close to 1/4, in which case multiple
topplings become possible. We even expect that for $\alpha<0.18$, this behavior becomes exact since most of the
avalanches are of size one. This result is in agreement with the  numerical results of \cite{Middleton}
 for $\alpha=0.07, 0.15$ and Eq.(\ref{mt}). For  $\alpha\ge 0.18$, it is given by Eq.(\ref{dissipationofc}) with
${\bar E}_i >1$ since the sites relax in (large) avalanches. This is confirmed numerically in Fig.\ref{frequency}.

\section{Synchronization and stochasticity}

\begin{figure}
\centerline{{\epsfxsize=7cm \epsffile{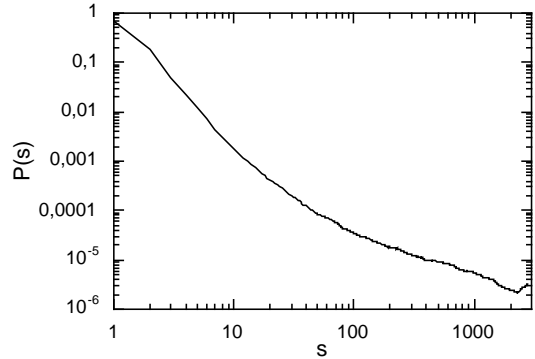}}}
\caption[Moyenne d'ensemble des distribution des tailles des avalanches pour le modle de Feder et Feder avec dŽsordre gelŽ]{\renewcommand{\baselinestretch}{1}\sl Average ensemble over 4000 realisations of the avalanche sizes of the disordered FF model for
$\alpha=0.2$ and $L=55$. For the avalanche of sizes in $[100,2800]$ the curve is  approximatively
$P(s)\sim s^{-0.8\pm0.04}$.}
\label{average}
\end{figure}

To study separately in a model of relaxation oscillators the tendency towards synchronization and the effect
of stochasticity, we have studied different versions of the
 Feder and  Feder (FF) model which is very close to the OFC model and where this separation is
possible\cite{FederI}. This model is identical to the OFC model but for the pulse which is a constant:
 $\Delta =\alpha E_c$. In this case, stochasticity of the initial conditions is almost completely dissipated
since the pulse is independent of the value of the oscillator before
 the relaxation. The only memory of the initial conditions lies in the hierarchy of values present initially
on the lattice and that determine in which order the oscillators relax. In fact, this model has the
unpleasant feature for us  of having, in the steady state, many sites exactly at the same value that therefore
 relax exactly at the same time, triggering disjoint avalanches. To remove this degeneracy that does not
occur in the OFC model and which is inconsistent with the slow drive limit -- the avalanches are supposed
instantaneous compared to the drive -- , we have added an infinitesimally small frozen disorder  $\delta_i$
on the thresholds $E_{ci}= E_c+ \delta_i$, $\delta_i\to 0$. In this case the analog of
Eq.(\ref{dissipationofc}) for the FF model is 
\begin{equation} 
({\bar E}_i- 4\alpha)r_i = 100\%
\label{dissipationff}
\end{equation} 
and the condition of stable synchronization analogous to Eq.(\ref{stressofc}) is
\begin{equation} 
({\bar E}_i- 4\alpha) {1\over 1-3\alpha} = 1\ \ \ \Longrightarrow\ \ \ {\bar E}_i=1+\alpha
\label{synchrostableff}
\end{equation} 
This condition is clearly non singular as $\alpha$ varies. Moreover, since there is almost no
stochasticity, we expect from the argument of Middleton and Tang that the system shows very stable
synchronization. This is indeed what we observe: in the steady state the system consists of (almost)
perfectly stable clusters of synchronized
 sites, the relaxation of which are triggered by sites near the borders. We have checked numerically that
Eq.(\ref{synchrostableff}) is effectively fulfilled and that the frequency of the synchronized clusters is
$(1-3\alpha)^{-1}$ thus proving that the mechanism of  Middleton and Tang works even for macroscopic
clusters of sites. A typical synchronized cluster is nucleated during the transient on the boundaries and
grows until it meets another cluster. Since both clusters have the same frequency of relaxation,
 they do not synchronize together in general. Since this process does not seem to 
involve any characteristic scale, we expect
that the distribution of these synchronized clusters follows a power law. By performing an ensemble average
over 4000 realisations -- obtained by choosing different initial conditions -- we have verified that, for
$\alpha=0.2$, the distribution of sizes of the synchronized clusters follows approximately for large
avalanches a power law, see Fig.\ref{average}. 
Moreover, we have verified that the size $s_{\mbox{\rm\scriptsize max}}$ of the largest
synchronized clusters vary with $L$ as $s_{\mbox{\rm\scriptsize max}}\sim L^2$, so that the synchronization of a macroscopic
fraction of the system should remain valid in the infinite volume limit. This behavior proves several
things. First, in the absence of stochasticity, the system is not critical. We have shown numerically that
it is in fact
 periodic in time  since  the same avalanches -- but for some sites at the edge of the avalanche -- occur
always at the same phase. Second, the mechanism proposed by Middleton and Tang works very well in this case
since the oscillators synchronize with the slowest ones even in macroscopic avalanches and the larger is
$\alpha$ the more rapid is the formation of the clusters, i.e. the more efficient is this mechanism.  Since
we know that in the OFC model there is only unstable synchronization even for relatively small $\alpha$ and 
SOC for $\alpha\ge0.18$ we can expect that stochasticity plays a role in destabilizing synchronization
and producing SOC. If this is true, the addition of a stochastic noise in the FF model should also produce
SOC. This is indeed what happens\cite{ChristensenI} and we have
 verified this fact for many values of $\alpha$ by adding to the pulse received by an oscillator $E_i$, a
very small stochastic noise $\Delta_i=\alpha E_i + \eta_i$ or equivalently by changing, after each relaxation,
the threshold $E_{ci}$ by a small stochastic
 amount, $E_{ci}\to E_{ci} + \eta_i$. In fact, it is possible to go smoothly from the disordered (frozen
disorder on the thresholds) to the noisy (stochastic noise) FF  model by updating the disorder on the
thresholds after each avalanche of only a fraction $q$ of the sites participating in the avalanche.
 For $q=0$, we get the disordered FF model while, for $q=1$, we get the noisy FF model. We have performed many
simulations and have observed that the system becomes
 critical already for rather small $q$, while for $q\to 0$ it is impossible to conclude since the system
stays stucked in quasi - stable synchronized situations for a very long
 time. However, we believe that even in this limit the system should be critical since the ensemble average,
previously mentioned for $q=0$, shows that the distribution of avalanche  sizes, which is frozen in each
realisation, follows a power law when averaged over different realisations. Therefore, for $q\ne 0$, the
stochasticity breaks the partially synchronized state into a partially and unstable synchronized one that
evolves more and more rapidly as $q$ approaches 1. We conjecture that the behavior of this noisy model
changes also at $\alpha_c$ since finite size scaling is obeyed also only above this value. Moreover, we have
checked  that the interior of the system is decoupled from the boundary for small values of $\alpha$, around
$\alpha=0.07$, and behaves exactly in the same way as in the OFC model. The level of synchronization
decreases of course as $q$ increases and for $q=1$ becomes very poor. We have also shown numerically that
for the same level of dissipation,
 the noisy FF model and the OFC model have the same exponent for the distribution 
of avalanche sizes, Fig.\ref{exponentofcff}.
This was done by direct comparison of the critical exponents that in both cases vary continuously with the
 dissipation and also by constructing a set of models that interpolate smoothly between these two models.

\begin{figure}
\centerline{{\epsfxsize=7cm \epsffile{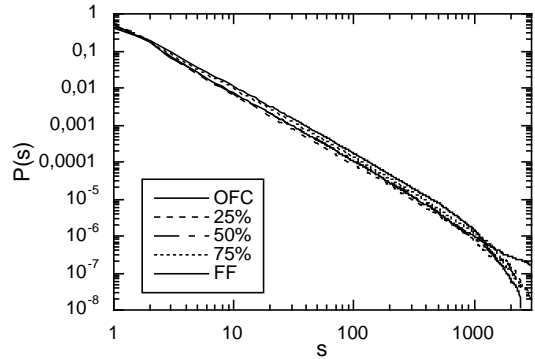}}}
\caption[Passage du modle d'Olami, Feder et Christensen au modle Feder 
et Feder]{\renewcommand{\baselinestretch}{1}\sl Distribution of the avalanche sizes for models 
that interpolate between the OFC and noisy FF models
for $\alpha=0.2$ and $L=90$. The different curves correspond to different percentages of sites that are
chosen of the OFC or FF type. }
\label{ofcff}
\end{figure}

This was done by building a series of models indexed by a probability $p$ such that after each avalanche the
sites that have just relaxed are chosen with probability $p$ to be for the next avalanche of the OFC type and
$1-p$ to be of the  noisy FF type. We have shown by varying $p$ between 0 and 1 that the slope
 of the distributions of the sizes (in a Log - Log plot) does not change, see Fig.\ref{ofcff}. Once again, the  OFC
models with $\alpha$ close to $1/4$ are particular cases since there is no conservative FF model and
therefore no such models in the same universality class.

Let us now come back to the OFC model. It is clear that this model shares a lot of properties with the noisy
FF model although the tendency towards synchronization and the level of stochasticity is controlled in the
OFC model by the same parameter $\alpha$.  We are now in a position to discuss our scenario about the
dynamics of the model. 

When $\alpha\le 0.18$, the dynamics is that described by Middleton and Tang. The
difference of frequencies between sites on the border and sites in the interior is not very large so that
the synchronized clusters are not very large. On the other hand stochasticity is weak so that the life time
of synchronized clusters is large. In the interior, the  sites are largely decoupled from the boundaries and relax
periodically with period $1-4\alpha$. 

 For $\alpha\ge 0.18$, the situation is rather different. The key
observation is that, contrary to the disordered FF model, the sites in the interior of the system cannot
synchronize in a stable way with sites on the border. This has two origins. First, stable synchronization
is in conflict with the randomness present in the model as in the case $\alpha\le 0.18$ and as in the
noisy FF model. Second, it is anyway impossible, for $\alpha>0.21$, to synchronize
 perfectly sites in the interior that have a natural frequency $(1-4\alpha)^{-1}$  with those on the
borders, the natural frequency of which is close to $(1-3 \alpha)^{-1}$,  because of the dissipation that
would not be $100\%$ in this case, if they were synchronized.
 However, nothing forbids a site  to be synchronized during a short time with a site on the boundaries, then
to be synchronized with another site of the boundaries and so on.  Thus,
 when $\alpha$ is large but not too close to 1/4, say $\alpha\sim 0.2$, neighboring sites in the interior
are still very well synchronized together and the level of synchronization decreases with the distance, see
Fig.\ref{mapsynchro}. This is exactly what is seen in the simulations. On the other hand, when  $\alpha\to 1/4$ the
synchronization time with a site on the boundaries goes to zero since the difference of frequencies between
these sites increases rapidly in this limit. Thus,  it is normal that the synchronization disappears in this
case. Once again, this is what is observed numerically even for values of $\alpha$ such as 0.24.

\begin{figure}
\centerline{{\epsfxsize=7cm \epsffile{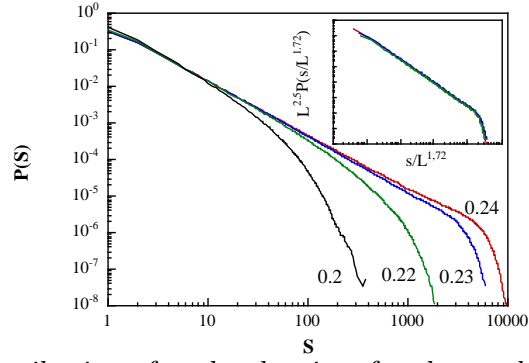}}}
\caption[Distribution des tailles des avalanches pour le modle d'Olami, Feder et Christensen avec voisins 
alŽatoires]{\renewcommand{\baselinestretch}{1}\sl Plot of the distribution of avalanche sizes for the 
random neighbor OFC model for 5600 sites. 
As shown in the inset for  $\alpha=0.23$, ordinary finite size scaling is
well verified since the curves obtained for $65^2,75^2$ and $95^2$ sites, can be superimposed  by a simple shift.}
\label{ofcrandom}
\end{figure}

 These results prove that
synchronization is not necessary for SOC in the OFC model. In fact, the example of the noisy FF model, which
is in the same universality class as the OFC model for $\alpha<1/4$ and that shows almost no synchronization,
was already an indication of this fact. Moreover, it is easy to build another model of relaxation
oscillators coupled {\it \`a la} OFC that, by construction, does not show synchronization: the random neighbor OFC
model in which a site that relaxes increments 4 sites chosen randomly. We have checked that this system is
critical for sufficiently large $\alpha$ but does not show  of course any synchronization, see Fig.\ref{ofcrandom}.

\section{Conclusion}
 In conclusion, we have shown
 that the OFC model can exhibit partial and unstable synchronization ($\alpha< 0.24$) with ($\alpha\ge
0.18$) or without ($\alpha\le 0.18$) being critical, and that it can be critical without showing
synchronization ($\alpha\ge 0.24$). While it is probable that in this system, the tendency towards
synchronization is indeed related to the mechanism that
 builds long range correlations, it is not necessary for criticality that the system actually shows
synchronization and it is even possible to build another model - the noisy FF model - that belongs to the
same universality class without showing synchronization for any value of $\alpha$. During the completion of
this work, we received an article of Lise and Jensen\cite{Lise} who performed the same simulation
 as us on the random neighbor OFC model and who also present a theoretical argument according to which the
model is critical only for $\alpha$ above 0.22, which is well verified numerically.  Although we agree with
their conclusion about the criticality of this model in the absence of synchronization, we disagree with the
fact that this random neighbor version of the model is a mean field approximation of the
 OFC model, or at least we believe that it is not proved. Strictly speaking, to be a mean field
approximation requires that this random neighbor version of the model shares the same critical exponents
with the original model above the critical dimension. This is not
 proved up to now, nor is it proved that in  dimensions higher than two, there exists a domain of $\alpha$ 
where SOC and synchronization coexist. Moreover, it is interesting to see that in fact both models -- at
least for the two dimensional OFC model -- are critical without synchronization for the same values of
$\alpha$. Therefore, we conclude that up to now there is no ``contradiction" between the behaviors of the
two models in this domain of parameters. Let us finally remark that the open boundary conditions in the
random neighbor model do not play the same role as in the OFC model and that there is not a unique way to
implement them. Therefore, as in the noisy FF model or the OFC model for $\alpha>0.21$, the random neighbor
OFC model is a SOC model of
 coupled relaxation oscillators that does not show synchronization, but up to now its relationship with the
OFC model is far from clear.

\medskip

\noindent Laboratoire de Physique Th\'eorique et Hautes Energies is a Unit\'e associ\'ee au CNRS: D 0280.

\medskip

\noindent  e-mail addresses: bottani@lpthe.jussieu.fr, delamotte@lpthe.jussieu.fr

\begin{thebibliography}{10}
\bibitem{note1}
For almost conservative systems, i.e. for $\alpha$ very close to $1/4$, the
  system needs multiple relaxations of the oscillators inside an avalanche to
  dissipate completely the stress it receives (see also
  \cite{Christensenqqch}).
\bibitem{note2}
This should not be confused with the change of behavior caused by the
  introduction of inhomogeneities in the model with periodic boundary
  conditions\cite{GrassbergerI,Torvund}.
\end{thebibliography}

\end{document}